\documentclass[fleqn,12pt]{wlscirep}
\usepackage[utf8]{inputenc}
\usepackage[T1]{fontenc}
\usepackage{newtxtext,newtxmath}
\usepackage{float}
\usepackage{gensymb}
\usepackage{soul}
\usepackage{enumitem}

\usepackage[numbers,sort&compress]{natbib}
\usepackage{setspace}
\usepackage{lineno}
\usepackage{float}
\title{Can AI weather models predict out-of-distribution gray swan tropical cyclones?}

\author[1]{Y. Qiang Sun}
\author[1,2]{Pedram Hassanzadeh}
\author[3]{Mohsen Zand}
\author[4]{Ashesh Chattopadhyay}
\author[5]{Jonathan Weare}
\author[1]{Dorian S. Abbot}
\affil[1]{University of Chicago, Department of the Geophysical Sciences, Chicago, 60637, IL}
\affil[2]{University of Chicago, Committee on Computational and Applied Mathematics, Chicago, 60637, IL}
\affil[3]{University of Chicago, Research Computing Center, Chicago, 60637, IL}
\affil[4]{University of California, Department of Applied Mathematics, Santa Cruz, 95064, CA}
\affil[5]{New York University, Courant Institute of Mathematical Sciences, New York, 10012, NY}
\affil[*]{qiangsun@uchicago.edu, pedramh@uchicago.edu}

\newpage
\setstcolor{red}
\begin{abstract}
Predicting gray swan weather extremes, which are possible but so rare that they are absent from the training dataset, is a major concern for AI weather models \textcolor{black}{and long-term climate emulators}. An important open question is whether AI models can extrapolate from weaker weather events present in the training set to stronger, unseen weather extremes. To test this, we train independent versions of the  AI \textcolor{black}{weather} model FourCastNet  on the 1979-2015 ERA5 dataset with all data, or with Category 3-5 tropical cyclones (TCs) removed, either globally or only over the North Atlantic or Western Pacific basin. We then test these versions of FourCastNet on 2018-2023 Category 5 TCs (gray swans). All versions yield similar accuracy for global weather, but the one trained without Category 3-5 TCs cannot accurately forecast Category 5 TCs, indicating that these models cannot extrapolate from weaker storms. The versions trained without Category 3-5 TCs in one basin show some skill forecasting Category 5 TCs in that basin, suggesting that FourCastNet can generalize across tropical basins. This is encouraging and surprising because regional information is implicitly encoded in inputs. Given that current state-of-the-art AI weather \textcolor{black}{and} climate models have similar learning strategies, we expect our findings to apply to other models. \textcolor{black}{Other types of weather extremes need to be similarly investigated.} Our work demonstrates that novel learning strategies are needed for AI models to reliably provide early warning or estimated statistics for the rarest, most impactful \textcolor{black}{TCs, and, possibly, other }weather extremes.

\end{abstract}

\begin{document}

\flushbottom

\maketitle

\thispagestyle{empty}
\newpage




\begin{doublespace}

\section*{Introduction}

Recent years have seen a rapid emergence of skillful artificial intelligence (AI) weather forecast models such as FourCastNet~\cite{pathak2022fourcastnet}, Pangu~\cite{bi2023accurate}, GraphCast~\cite{lam2022graphcast}, AIFS~\cite{lang2024aifs}, Fuxi~\cite{chen2023fuxi}, and Stormer~\cite{nguyen2024scaling}. These data-driven models are deep neural networks (NNs) that predict the evolution of the 3D global atmospheric state in six or twelve hour increments after being trained on the ERA5 reanalysis dataset from 1979-2015. AI weather models' out-of-sample (2018-2023) forecasts of the global weather, including some aspects of extreme events such as the track of tropical cyclones (TCs), have been shown to outperform predictions from the best numerical models for up to 10 days~\cite{bi2023accurate, lam2022graphcast,   rasp2024weatherbench, ben2024rise}. Aside from increased accuracy, a major advantage of AI weather models is that, once trained, they can be run $10^4$--$10^5$ times faster than state-of-the-art numerical models~\cite{kurth2023fourcastnet,bi2023accurate,lam2022graphcast}, and for example, produce large-ensemble and probabilistic forecasts~\cite{price2023gencast,li2024generative,mahesh2024huge, mahesh2024huge2}. Furthermore, this substantial speed-up has opened a new, rapidly advancing avenue for developing AI \textcolor{black}{models for long-term emulation of the atmosphere, ocean, or the entire climate system (we refer to such models as AI climate emulators, hereafter) } ~\cite{watt2023ace,cachay2024probabilistic, duncan2024application,guan2024lucie,lai2024machine,cresswell2024deep,dheeshjith2024transfer,bracco2024machine,kochkov2024neural}.
The speed of such emulators\textcolor{black}{, which can be trained on reanalysis datasets and/or global climate model (GCM) outputs,} enables the generation of large ensembles of long runs. \textcolor{black}{It has been suggested that such ensembles could} significantly reduce sampling error (i.e., the internal variability uncertainty), a major challenge for climate change projections of extreme weather events, particularly at regional scales \cite{lai2024machine,chattopadhyay2023long, cresswell2024deep,bracco2024machine}.

A key question relevant to the fidelity and usefulness of AI weather \textcolor{black}{models} and climate emulators is their ability to forecast/emulate the rarest, yet most impactful, extreme events~\cite{field2012managing, ragone2018computation,eyring2024pushing}. The rarity of the most extreme weather events makes them harder to learn for AI models, which is the classic ``data imbalance'' problem in statistical learning~\cite{krawczyk2016learning, chattopadhyay2019analog, walsh2020extreme, ebert2020evaluation, miloshevich2023probabilistic}. AI weather models have often been found to underestimate the peak amplitude of events such as heat waves and TCs~\cite{pasche2024validating, charlton2024ai,demaria2024evaluation}, which may result from data imbalance as well as other \textcolor{black}{problems} such as blurring due to spectral bias \cite{chattopadhyay2023long,bonavita2024some,selz2023can}. AI weather models have shown remarkable skill predicting TC tracks (much better than TC intensity)~\cite{pathak2022fourcastnet,bi2023accurate,lam2022graphcast,demaria2024evaluation}; however, TC tracks are largely determined by large-scale background winds that are not rare.



Data imbalance reaches its asymptotic limit for so-called ``gray swan'' weather extremes, first defined by Lin and Emmanuel~\cite{lin2016grey} as rare weather events that are physically possible but have never actually been observed in the historical record. In the context of AI models in climate science, gray swans are physically possible weather events that are rarer and stronger than those in the training set, thus they are ``out of distribution''.  While in general, \textcolor{black}{neural network-based} AI algorithms should not be expected to forecast/emulate out-of-distribution events, one might hypothesize that for physical systems such as climate, AI models could learn key physical relationships among variables from weaker (in-distribution) events that would generalize to stronger (out-of-distribution) events. Although this hypothesis has not been rigorously tested for extreme weather events with state-of-the-art AI models, there are recent studies that suggest these AI weather models learn dynamics to some degree, rather than simply memorizing patterns. These include the interesting results of Hakim and Masanam~\cite{hakim2024dynamical}, who showed the ability of AI weather models to predict the response of the atmosphere to highly localized (but not extreme) perturbations that are substantially different from any pattern in the training set, and the work of Rackow \textit{et al.}~\cite{rackow2024robustness}, who demonstrated these models' relatively robust short-term forecasting skill under climate change. 

\begin{figure}[htbp]
\centering
\includegraphics[width=\linewidth]{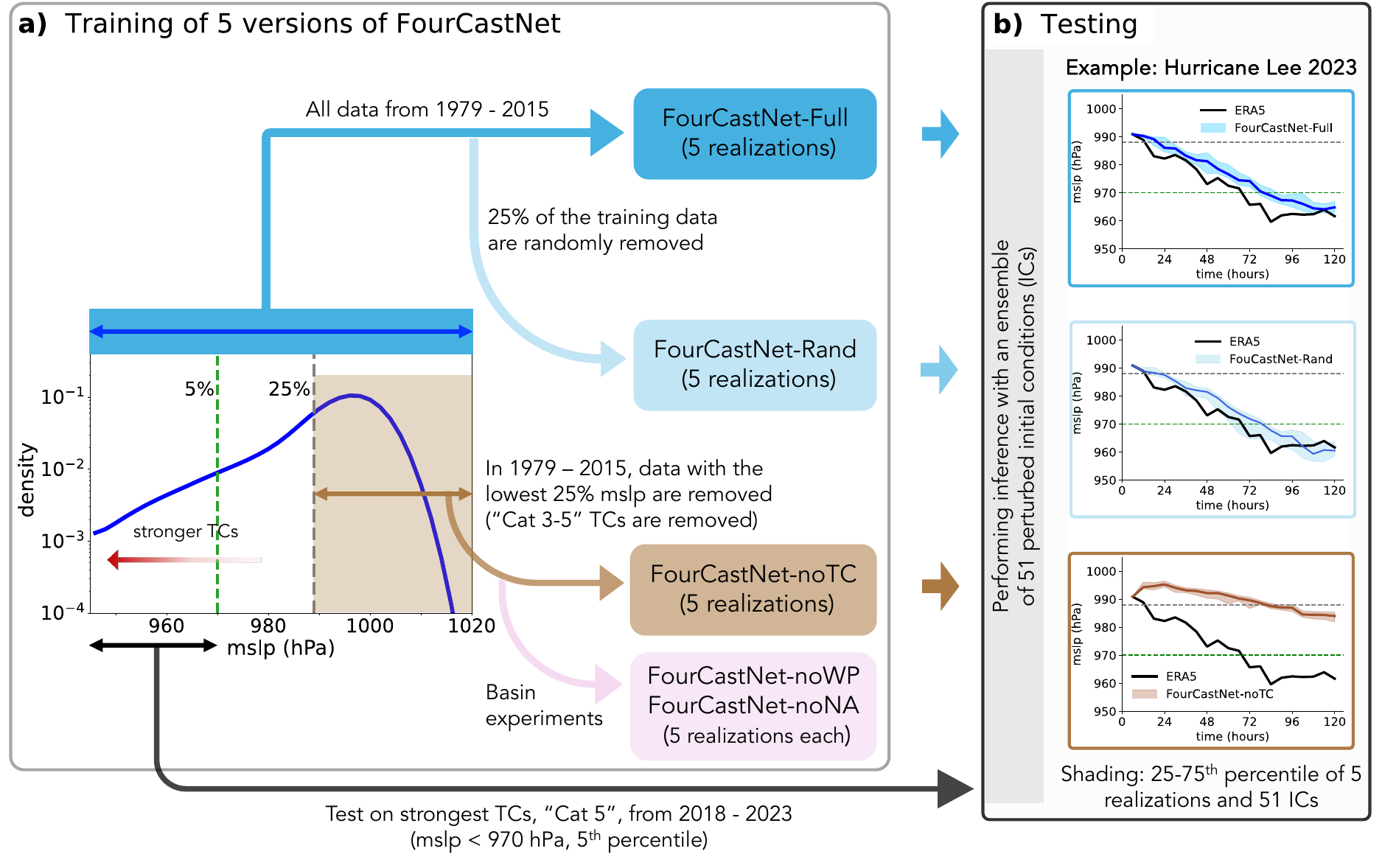}
\caption{\textbf{Schematic overview of this study.} \textbf{a) Training of five versions of FourCastNet.} The panel depicts the histogram of minimum mslp in the tropics (30°S–30°N) in the training set (ERA5, 1979-2015). Note that a lower mslp corresponds to a stronger TC. Vertical lines indicate the 5th and 25th percentiles, which are 970 and 988~hPa, respectively. For FourCastNet-Full, the full training dataset is utilized. For FourCastNet-noTC, samples with instances of mslp below $988.0$~hPa anywhere in the tropics are removed from the training set. FourCastNet-Rand uses a training set of the same size and seasonal distribution as noTC but with samples removed randomly (while ensuring that samples with mslp~$<988.0$~hPa are retained). Two additional models are also trained for which samples below 988~hPa only over the tropical Western Pacific (noWP) or tropical North Atlantic (noNA) basin are removed. For each training set, five independent versions (realizations) are trained from different random weight/bias initializations to account for model uncertainty. \textbf{b) Testing of the five models.} The forecast skill of each trained model is evaluated for TCs with mslp below 970~hPa (Category 5) in the test set. The right panels provide an example of the forecast results for Hurricane Lee (2023), a Category 5 TC. Shading represents the 25th to 75th percentile range of forecasts, derived from five model realizations and 51 different initial conditions (ICs) provided by an ensemble of data assimilations (EDA) from ECMWF; See Methods and Data.}
\label{fig:schematic}
\end{figure}

The main objective of this paper is to examine, in highly controlled experiments, whether the AI weather model FourCastNet can forecast gray swan TCs. Our framework is summarized in Fig.~\ref{fig:schematic}. Briefly, we have created 4 additional training sets from the original ERA5 1979-2015 training set (Full). In one set (noTC), we have removed any training sample that contained mean sea-level pressure (mslp), a measure of TC strength (the lower mslp, the stronger TC), below the 25th percentile (988~hPa) in the tropics (30S-30N); see Methods and Data. This is roughly equivalent to removing samples containing TCs of Category 3-5 anywhere in the world. Our next training set (Rand) has the same size and seasonal distribution as noTC, but we have randomly removed $25\%$ of the samples from the full set, while avoiding removing any Category 3-5 TCs (see Methods and Data). Finally, we have created two additional training sets by removing samples containing tropical mslp below 988~hPa over either the Western Pacific (noWP) or North Atlantic (noNA) basins. We then train 5 versions of FourCastNet on each training set from 5 different random realizations of the weights and biases, leading to 25 independently trained models. Next, we quantify the forecast skill of each model on the mslp evolution of the strongest (lowest 5th percentile) TCs, roughly corresponding to Category 5, in the test period (2018-2023). 

Before presenting results in the next section, we emphasize that while here we focus on one AI model (due to the computational cost of training) and one type of extreme event\textcolor{black}{s} (TCs), \textcolor{black}{we speculate that} the findings and insights of this work are likely to apply to other current state-of-the-art AI weather/climate models, which share the same principal learning process\textcolor{black}{. This point and the potential implications for} other kinds of extreme events \textcolor{black}{are discussed further in} Summary and Discussion.

\newpage

\section*{Results}

\subsection*{Failure to Extrapolate to Out-of-Distribution Gray Swan TCs}

Figure~\ref{fig:schematic}(b) compares the forecasting skill of FourCastNet-Full, -Rand, and -noTC on Category 5 Hurricane Lee (2023). With its winds increasing by 85~mph (140~km/h) in 24 hours, Hurricane Lee underwent rapid intensification before reaching its peak intensity. TCs of similar intensity to Hurricane Lee are present in the training datasets of FourCastNet-Full and FourCastNet-Rand. Both models are able to forecast Lee's rapid intensification fairly well, although their forecasted mslp evolution has an approximately one-day lag relative to ERA5. Most importantly, almost all 255 ensemble forecasts with these two models reach a minimum mslp below 970~hPa, the threshold for Category 5 TCs in ERA5 data (see Methods and Data). In contrast, there are no comparably strong TCs in the noTC training dataset (mslp $\ge$ 988 hPa), so Hurricane Lee represents an out-of-distribution gray swan event for FourCastNet-noTC. The performance of FourCastNet-noTC on Hurricane Lee is much worse than that of FourCastNet-Full and FourCastNet-Rand, suggesting difficulty predicting out-of-distribution events. In particular, all members of the FourCastNet-noTC forecast a weakening of Lee (i.e., mslp increases) from the beginning (\ref{fig:schematic}b), followed by a very slow re-intensification, such that their mslp stays above 980~hPa during the 5-day forecast period. We note here that the mslp threshold for noTC was 988~hPa, which means that FourCastNet-noTC's predictions barely go below the lowest mslp seen in the tropics. Thus, FourCastNet-noTC is not only inaccurate, but its forecasts yield ``false negatives,'' the worst type of error for decision-critical tasks: The model forecasts a moderate Category 3 TC rather than a devastating Category 5 TC, giving no signal that its forecast is inaccurate. As we show below, this is a consistent behavior of FourCastNet-noTC. 

\begin{figure}[htbp]
\centering
\includegraphics[width=\linewidth]{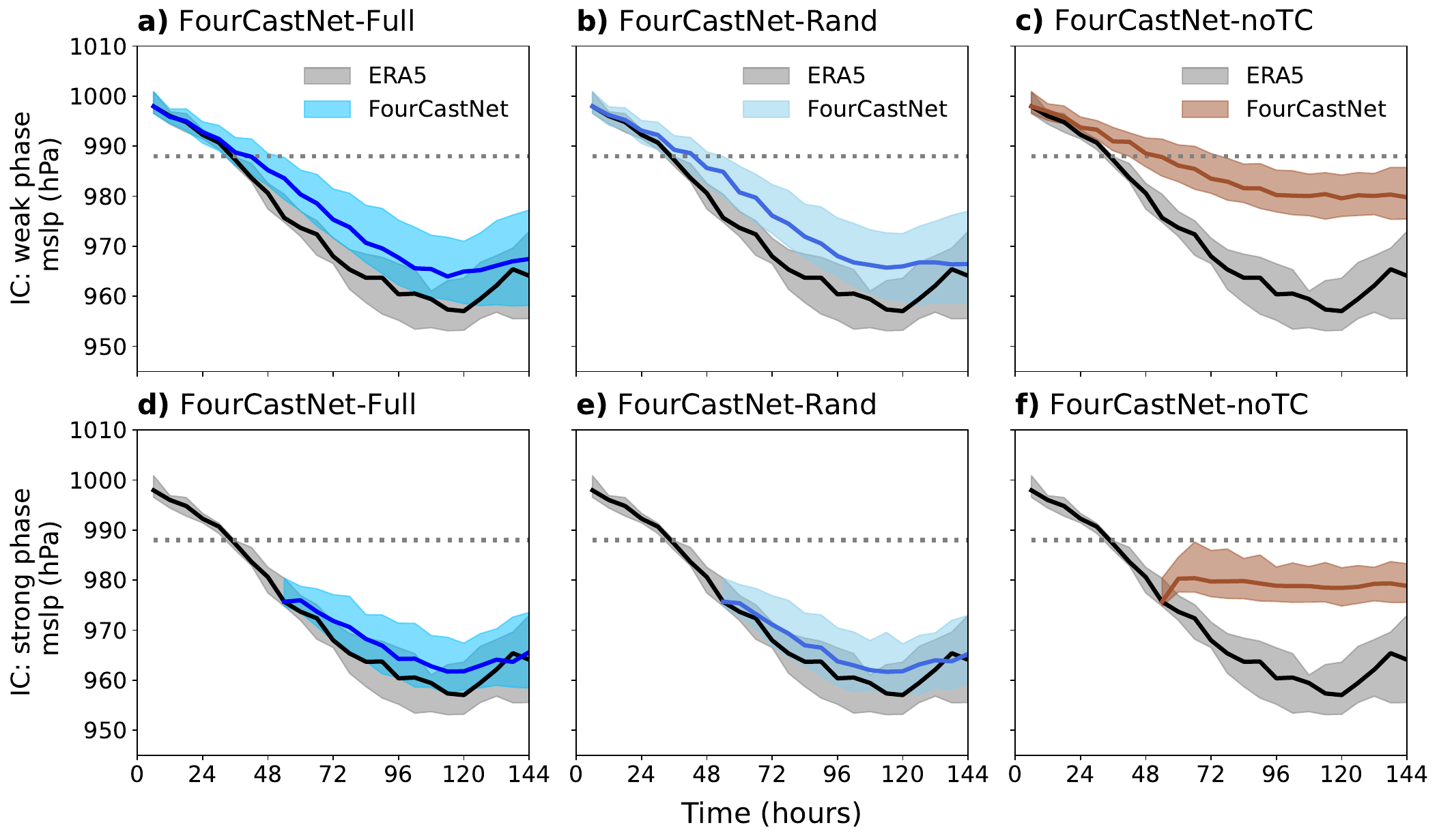}
\caption{\textbf{FourCastNet's difficulty in extrapolating to gray swan TCs.} Forecasting of all 20 Category 5 TCs from the test set (2018-2023) by three versions of FourCastNet trained on different datasets: FourCastNet-Full (left column), FourCastNet-Rand (middle column), and FourCastNet-noTC (left column). Dashed line shows the critical threshold for 25th percentile of minimum mslp (roughly Category 3 TC) used in the noTC training set. All panels show the evolution of the median mslp (solid line) and the inter-quartile range from the 25 to the 75th percentile (shading) over all 20 Category 5 TCs, 5 realizations of each trained model, and 51 perturbed initial conditions from EDA (5100 forecasts). Shading for ERA5 is over the 20 TCs. Forecasts are initialized one day before each TC reached the critical threshold (weak phase, top row) or one day after the TC reached this threshold (strong phase, bottom row). The latter initial conditions are out-of-distribution. \textcolor{black}{As an additional note, detailed analysis shows that none of the ensemble members in the FoureCastNet-noTC forecasts reached the observed lowest mslp values. Although a few members' mslp reached 970 hPa, this occurred because these members transitioned to an unstable state that eventually led to blow-up, rather than capturing realistic intensification of the storm.}}
\label{fig:full-rand-notc}
\end{figure}

Figure~\ref{fig:full-rand-notc} presents similar comparisons, but aggregated over all 20 Category 5 TCs in the testing dataset, and includes forecasts initialized in \textcolor{black}{both the weak and strong phases} of the TCs - during the transition from Category 4 to 5 (second row). The conclusions are the same as those drawn for Hurricane Lee. Both FourCastNet-Full and FourCastNet-Rand perform fairly well in forecasting the evolution of minimum mslp, whether the initial conditions are during the weak or strong phase of the TCs. They do show a consistent bias in underpredicting the strength of TCs, similar to results reported for other state-of-the-art AI weather models for TCs \cite{demaria2024evaluation} and other extreme events \cite{charlton2024ai}. This bias could result from data imbalance or other problems such as blurring due to spectral bias \cite{chattopadhyay2023long,bonavita2024some}. 

In contrast, FourCastNet-noTC significantly underpredicts the evolution of mslp (panels (c) and (f)), \textcolor{black}{despite demonstrating comparable performance to FourCastNet-Full in predicting the tracks of TCs (Figure S1).} Although FourCastNet-noTC has some skill on the first day of intensification when initialized in the  weak phase, once the ERA5 mslp approaches the critical value of 988~hPa after around one day, the intensification in FourCastNet-noTC's forecasts slows dramatically and mslp barely reaches below $980$~hPa. When FourCastNet-noTC is initialized in the strong phase, where the initial conditions are out-of-distribution, FourCastNet-noTC forecasts a weakening of the TC. It appears that the model tries to relax toward the values of intensity it had seen in training (above $988$~hPa), instead of intensifying to the lower values of mslp observed in ERA5. As noted above, this behavior would lead to false negative forecasts for destructive storms, which has major societal consequences.  Additionally, it is important to note that the similar performance of FourCastNet-Full and -Rand on Category 5 TCs shows that a $25\%$ shorter training set does not impact the forecasting skill, so this cannot be the reason for FourCastNet-noTC's poor performance. 


These comparisons consistently show that FourCastNet is unable to learn from weaker TCs (Category 1-2) and extrapolate to stronger, unseen Category 5 TCs. Despite an inability to extrapolate to gray swan TCs, FourCastNet-noTC performs similarly to FourCastNet-Full and FourCastNet-Rand on typical weather that all models see in their training set. Specifically, \textcolor{black}{the three models exhibit similar forecast skill for global and tropical weather (based on anomaly correlation coefficient (ACC) and root-mean-square error (RMSE)), as shown in Figure~S2, and for weaker (Category 1-2) TCs in the testing set, as presented in Figure~S3.}






\newpage

\subsection*{Some Success Generalizing Across Tropical Basins}

\begin{figure}[htbp]
\centering
\includegraphics[width=\linewidth]{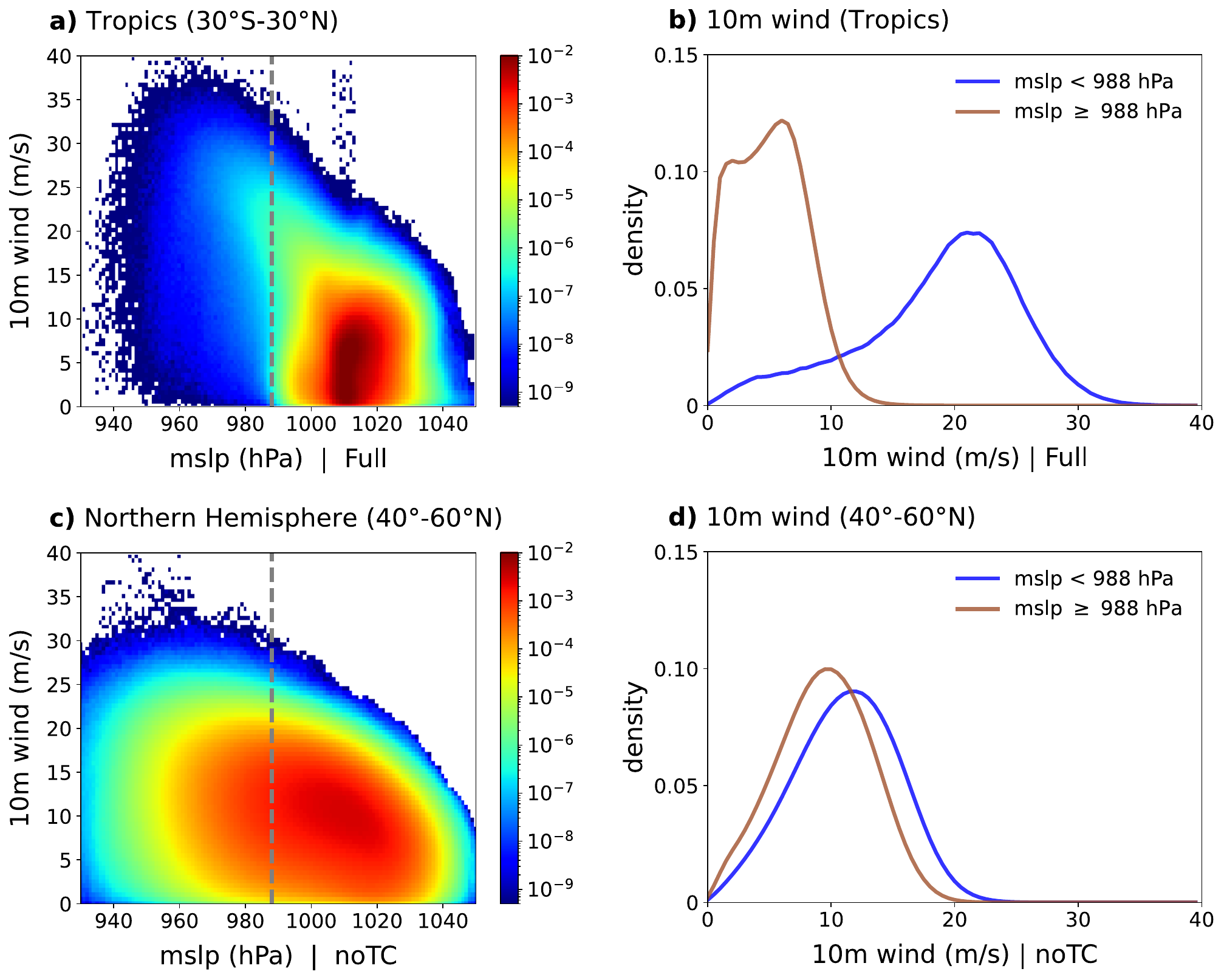}
\caption{\textbf{Extratropical cyclones and TCs exhibit different dynamical behavior.} \textbf{a}) Joint PDF of mslp and 10-meter winds in the tropics (30°S–30°N) in the Full training set. \textbf{b}) Probability density of 10-meter winds in the tropics in the Full training set, conditioned on the mslp threshold. \textbf{c}) Similar to (a), but for the midlatitudes (40°–60°N) of the noTC training set. \textbf{d}) Similar to (b), but for the midlatitudes of the noTC training set.}
\label{fig:mid-latitdue-tropics-pdf}
\end{figure}

Although the training dataset of FourCastNet-noTC has no TC (or any sample) with mslp below $988$~hPa in the tropics, it has plenty of samples with mslp below this value in the midlatitudes, associated with extratropical cyclones. This can be seen in Figure~\ref{fig:mid-latitdue-tropics-pdf}(a) and (c) and \textcolor{black}{Figure~S4(a)}, which show that accounting for the values of mslp globally in the noTC training set, mslp $< 988$~hPa is not unprecedented or out-of-distribution at all. 

One might wonder why \textcolor{black}{the presence of strong extratropical cyclones in the training set does not enable} FourCastNet-noTC to \textcolor{black}{predict} Category-5 TCs with similarly low mslp. We believe that this is due to the difference in the dynamics of TCs and extratropical cyclones. TCs' dynamics are driven by convection and latent heating with azimuthal winds reaching the maximum near the low-pressure center, while extratropical \textcolor{black}{cyclones}' dynamics are driven by baroclinic instability and their wind profiles are different. For the AI model, these differences are manifested in the evolution of the entire state vector (input) rather than a single variable (like mslp), and can be seen, for example, in the joint PDFs in Figure~\ref{fig:mid-latitdue-tropics-pdf}. While there is a marked trend of increases in wind speed as mslp decreases below 1000~hPa in the tropics, in the extratropics, the relationship between mslp and wind speed is much less pronounced, such that it is not uncommon to have low wind speeds when the mslp is low. The distinction between the two phenomena can be also seen in the probability density of 10-meter wind for weak (mslp $>988$~hPa) and strong (mslp $<988$~hPa) cyclones. The different physical mechanisms driving extratropical cyclones and TCs therefore lead to different relationships between physical variables in space and time. \textcolor{black}{As a result, the presence of strong, low-mslp extratropical cyclones (in the training set) does not help FourCastNet-noTC with predicting Category 5 TCs in the tropics during testing.}



\begin{figure}[htbp]
\centering
\includegraphics[width=0.8\linewidth]{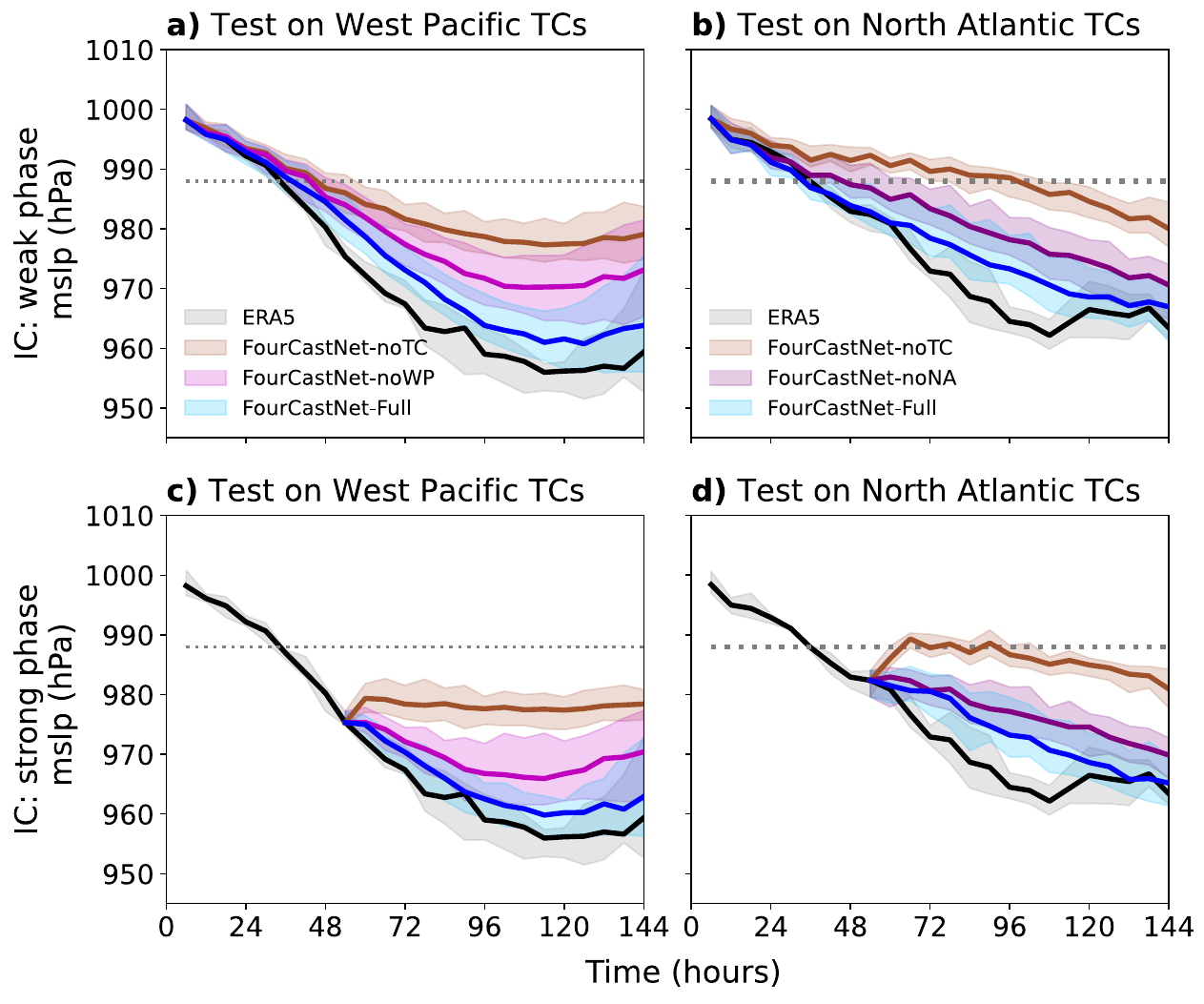}
\caption{\textbf{FourCastNet generalizes across tropical regions for dynamically similar events.} \textbf{a}) Comparison of the forecast skill of FourCastNet-noWP against other models for Category 5 TCs (from the test set) in the Western Pacific, initialized at the TC's weak phase. \textbf{b}) As in (a), but for TCs in the North Atlantic basin. \textbf{c})-(\textbf{d}) As in (a)-(b), but initialized at the strong phase of the TCs. Solid lines and shading are as in Figure~\ref{fig:full-rand-notc}.}
\label{fig:basin-results}
\end{figure}

Given the above discussion, one might next wonder whether FourCastNet can generalize from strong TCs it has seen in the training set in one tropical basin to another. While there are differences between the TCs in the two major basins of activity, the North Atlantic and Western Pacific, mainly due to differences in large-sale circulation, TCs in both regions are fundamentally driven by similar dynamical processes. Substantial similarity between the mslp and 10-meter wind distributions of TCs in these two basins \textcolor{black}{(Figures S5(e-h))}. Both FourCastNet-noWP and FourCastNet-noNA perform significantly better than FourCastNet-noTC when tested on Category 5 TCs in the Western Pacific and North Atlantic basins, respectively (Figure~\ref{fig:basin-results}). Specifically, \textcolor{black}{FourCastNet-noWP and FourCastNet-noNA} forecast intensification of all tested TCs from both weak and strong initial conditions and produce minimum mslp well below 970~hPa for many storms (\textcolor{black}{Figure~\ref{fig:basin-results}}). In fact, FourCastNet-noWP's performance is similar to that of FourCastNet-Full for 3 out of the 13 tested TCs in the WP basin (not shown). The reduced performance of FourCastNet-noWP and FourCastNet-noNA relative to FourCastNet-Full may be due to the smaller number of intense TCs in their training datasets. 

Overall, this analysis suggests that FourCastNet can effectively generalize across geographic regions by learning from dynamically similar events. We emphasize the need for dynamical similarity given that FourCastNet-noTC could not generalize from strong extratropical cyclones to unseen strong TCs.



\newpage

\newpage
\subsection*{Forecasts Lack Physical Consistency}
Adding physical constraints is often cited as an avenue for improving the skill of AI weather/climate models for extreme weather events and out-of-distribution generalization~\cite{charlton2024ai,lai2024machine,bracco2024machine,beucler2024climate}. Here, we examine a key physical balance of TCs in all trained versions of FourCastNet to see whether the poor out-of-distribution generalization to gray swans in FourCastNet-noTC could be due to lack of this balance. 
Above the boundary layer, TCs approximately satisfy the gradient-wind balance between the pressure gradient force and the Coriolis and centrifugal forces in the azimuthal mean \citep{willoughby1990gradient}:
\begin{equation}
    {g} \frac{\partial Z}{\partial r} = \frac{V_g^2}{r} + f V_g.
    \label{eq:GradWind}
\end{equation}
Here, $V_g$ is the azimuthal gradient wind, $Z$ is the geopotential height (the height of a given pressure surface), $g$ is the gravitational acceleration, $r$ is the radial distance from the center of the cyclone, and $f$ is the Coriolis parameter.

\begin{figure}[htbp]
\centering
\includegraphics[width=\linewidth]{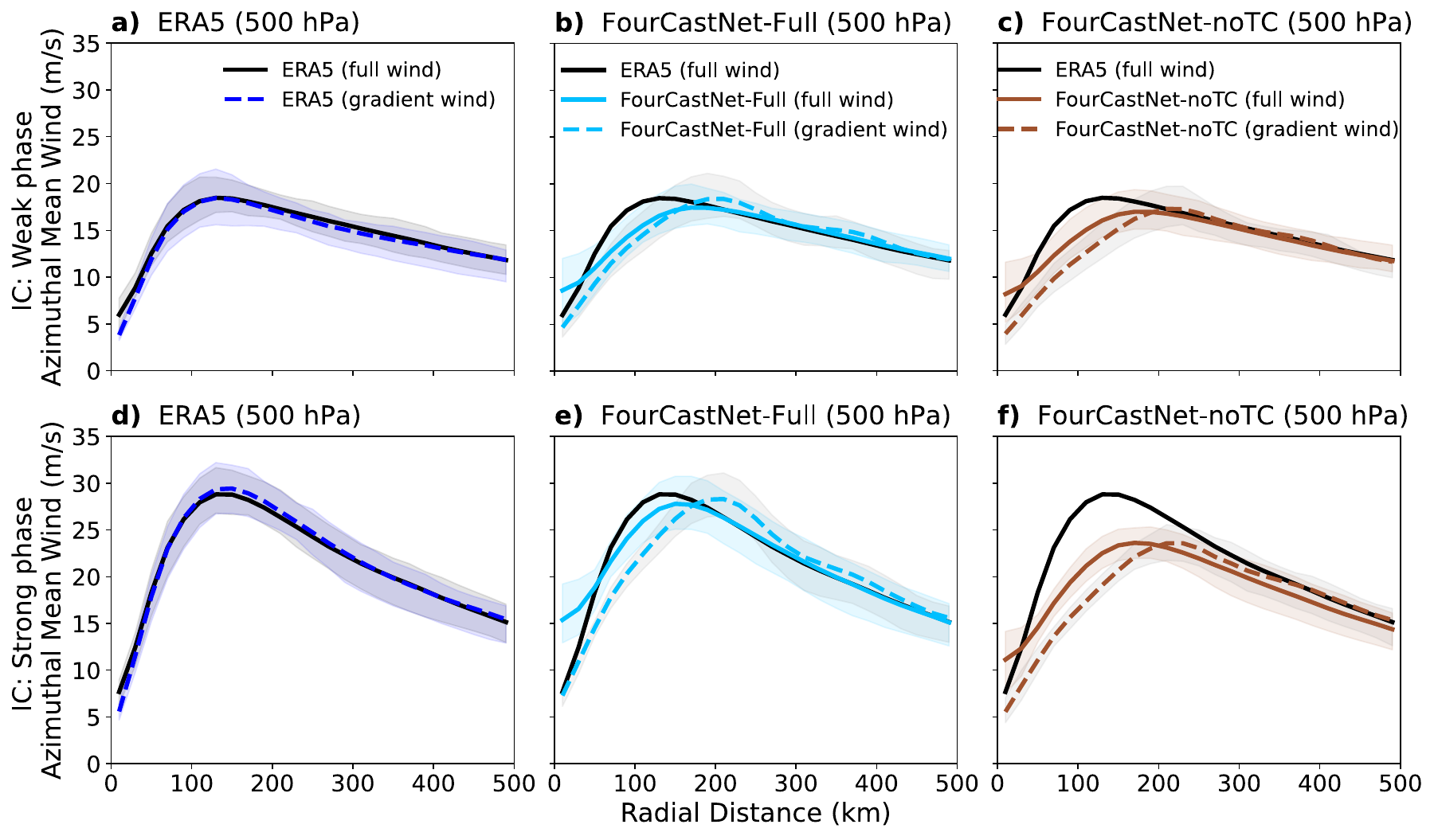}
\caption{\textbf{Lack of physical consistency in the forecasts.} \textbf{a) Gradient-wind balance in ERA5}. Radial profiles of azimuthal wind and the gradient-wind derived from  Eq.~\eqref{eq:GradWind} at 500~hPa for all Category 5 TCs in the test set (2018-2023) in their weak phase. \textbf{b}) As in (a), but for FourCastNet-Full's forecasts. (\textbf{c}) As in (b), but for FourCastNet-noTC's forecasts. The bottom row is the same as the top row, but for the strong phase of Category 5 TCs. The shading indicates the 25th to 75th percentile range across the 20 Category 5 TCs in the test (left panel). In the middle and right panels, the shading is over the 20 TCs, 5 realizations, and 51 perturbed ICs.}
\label{fig:gradient-wind-balance}
\end{figure}

As expected, the gradient-wind balance holds for Category 5 TCs in the ERA5 data across all radial length scales (Figure~\ref{fig:gradient-wind-balance}). This reflects the fact that ERA5 is based on a physical model with relatively small adjustments during data assimilation to better fit with observations, and the dominant balance in the equations of motion for TCs is gradient-wind balance (Eq.~(\ref{eq:GradWind})). In contrast, the gradient-wind balance is not observed within $\sim 200$~km from the center of TCs in forecasts from both FourCastNet-Full and FourCastNet-noTC. \textcolor{black}{The most significant deviations from balance, i.e., the difference between the full wind and the gradient wind, occur when forecasts are initialized from the strong phase (Figures 5(e), (f)).  While the magnitude of the gradient wind is comparable to the full wind, the radius of maximum gradient wind is too large relative to that of the full wind. Although FourCastNet-Full appears to simulate gradient wind balance slightly better than FourCastNet-noTC when forecasts are initialized from the weak phase (Figures 5b, c), the difference is within the uncertainty range.}  Interestingly, despite the fact that FourCastNet-Full is much more successful at forecasting the evolution of Category 5 TCs than FourCastNet-noTC, the wind and pressure fields are no more physically consistent in FourCastNet-Full than in FourCastNet-noTC.

Lack of physical consistency is not unique to FourCastNet. A few recent studies have also pointed out lack of physical balances in other state-of-the-art AI weather models \cite{bonavita2024some,charlton2024ai}. The implications of these findings for potential avenues for improving the out-of-distribution generalization of AI weather/climate models will be discussed in the next section.

\newpage

\section*{Summary and Discussion}
We conduct controlled experiments in which we train the AI weather model FourCastNet after removing Category 3-5 TCs from the training sets, either globally (FourCastNet-noTC) or only over the Northern Hemisphere Atlantic or Pacific basins (FourCastNet-noNA and -noWP). By analyzing the forecasting skill of these different models on Category 5 TCs in the test set we demonstrate the following:
\begin{enumerate}
    \item \textbf{Lack of out-of-distribution generalization (extrapolation) for \textcolor{black}{gray swan TCs}:} FourCastNet is \textit{unable} to learn about unseen strong TCs (Category 5) from the weaker ones (Category 1-2) that were present in the training set (Figures~\ref{fig:schematic}-\ref{fig:full-rand-notc}). \textcolor{black}{This is despite the fact that low mslp extratropical cyclones (at the level of Category 5 TCs) exist in the  noTC training set. However, because extratropical cyclones and TCs have different dynamics (as, for example, manifested in the joint PDFs of mslp-wind, Figures~\ref{fig:mid-latitdue-tropics-pdf} and S4), the low mslp values in extratropical cyclones could not help FourCastNet-noTC with predicting Category 5 TCs}. 
    \item \textbf{Some generalization for dynamically similar \textcolor{black}{storms} across ocean basins:} FourCastNet is \textit{able} to learn something about unseen strong TCs in one ocean basin from strong TCs it has seen in another basin in the training set (Figure~\ref{fig:basin-results}). This is somewhat surprising, \textcolor{black}{but encouraging,} as \textcolor{black}{location-specific} information (e.g., latitude, longitude, topography) \textcolor{black}{is} implicitly encoded in the inputs. 
    \item \textcolor{black}{\textbf{Lack of gradient-wind balance:}} Whether trained with the Full or noTC dataset, FourCastNet is \textit{unable} to reproduce gradient-wind balance, a key physical constraint that is obeyed by TCs in the training set (Figure~\ref{fig:gradient-wind-balance}). \textcolor{black}{Enforcing gradient-wind balance in the loss function could potentially help the model learn this physical relationship, though there could be tradeoffs in the accuracy or the representation of other physical properties (e.g., the ageostrophic flow)}.
    \item \textbf{Common metrics can obscure poor performance for gray swan \textcolor{black}{TCs}:} FourCastNet-Full, FourCastNet-Rand, and FourCastNet-noTC show similar forecast skill as measured by \textcolor{black}{ACC or RMSE calculated both globally and over the tropics, as well as by forecasting skill on in-distribution TCs (Category 1-2 here); see~Figures~S1-S3}.
\end{enumerate}

As discussed below, (1)-(2), which have been unambiguously tested for an AI weather model for the first time here, along with (3)-(4), have major implications for efforts to understand and predict \textcolor{black}{TCs using AI weather models. We speculate that our findings might have implications for the gray swans of other types of extreme events and for AI climate emulators.} 

Before discussing these implications, potential solutions, and proper tests for learning gray swans, we highlight that here we have focused on only one model, FourCastNet. We did this because it was the first state-of-the-art AI weather model publicly available for training and is \textcolor{black}{much  cheaper to train than new models like PanguWeather or GraphCast~\cite{bi2023accurate,lam2022graphcast,guo2024fourcastnext}}. Still, training 25 independent versions was computationally demanding (note that in this work, each version had to be trained from random initial weights; we cannot finetune a pre-trained FourCastNet for our experiments). However, given that current state-of-the-art AI weather models and climate emulators share the same principal learning process (i.e., physics-free deterministic or probabilistic evolution of the mapping of $\mathbf{x}(t)$ to $\mathbf{x}(t+\Delta t)$), we expect limitation (1) to apply to them as well. In fact, while newer models often show improved accuracy \textcolor{black}{on global and some regional metrics}, recent studies have found similar types of shortcomings, e.g., physical inconsistency like (3) or missing the peak amplitude of extreme events, in different state-of-the-art models~\cite{bonavita2024some,chattopadhyay2023long, charlton2024ai} \textcolor{black}{(note that \cite{demaria2024evaluation} found FourCastNet to have forecast skill for TCs comparable to newer models)}. Even most of the emerging ``foundation weather/climate models''~\cite{nguyen2023climax,schmude2024prithvi,wang2024orbit}  use the same overall principal learning process. There are recent examples of AI models that use self-supervised learning algorithms, such as pre-training with masked-autoencoders~\cite{man2023w,lessig2023atmorep,mcnally2024data}, \textcolor{black}{or hybrid models such as NeuralGCM~\cite{kochkov2024neural}, although, at least so far, these approaches do not have any component to address data imbalance.} Whether they improve (1) and (3) remains to be thoroughly investigated, and should not be assumed without rigorous demonstration (see below).

We have focused here on only one type of extreme weather (TCs), again due to computational cost. Whether (1)-(2) apply to other major types of extreme weather events needs to be \textcolor{black}{thoroughly} studied using similar controlled experiments. While stronger TCs could not be learned from the weaker ones, it is possible that some types of extreme events could be learned from weaker examples. Furthermore, extreme weather events often have distinct dynamics (e.g., dry and moist heat waves, atmospheric rivers, and cold snaps), which can hinder generalization among them~\textcolor{black}{.} However, some of the main physical processes, such as zonal and meridional thermal advection in heat and cold waves, share similarities. To what degree learning one type of event can translate into another type remains to be seen. Our \textcolor{black}{current} results clearly demonstrate that the presence of other related extreme events (i.e., extratropical cyclones) does not help with \textcolor{black}{gray swan} TCs. 

\textcolor{black}{The main \textcolor{black}{potential} implications of (1)-(4) for current state-of-the-art AI weather models and climate emulators, are discussed below. We emphasize that these items are of significant importance but are currently speculative and require extensive and careful investigation.} 
\begin{enumerate}
  \item[a.] \textbf{AI weather models \textcolor{black}{might} produce unreliable early warning for \textcolor{black}{some types of} unprecedented weather events:} AI weather models \textcolor{black}{might} fail to accurately forecast weather extremes that are \textit{unprecedented globally}. However, it is encouraging that if dynamically similar events of comparable or larger amplitudes exist in other parts of the world in the training set, the AI models may show some forecast skill. 
  \textit{The possibility of AI weather models misforecasting extreme weather events, particularly if they produce the ``false negatives'' we consistently find here, creates serious societal risk}. This concern is particularly acute as climate change is increasing the likelihood of some types of gray swans (including TCs), just as rapid advances in AI weather forecasting invite more operational reliance on such models over traditional physics-based ones. This suggests that the limitations of the AI weather models for forecasting gray swans must be fully characterized (see (c)). Additionally, it would be valuable to identify early warning signals that an AI \textcolor{black}{weather} model is failing. \textcolor{black}{As highlighted above}, \textcolor{black}{the strengths and weaknesses of AI models for gray swans might vary among different types of extreme weather events, requiring comprehensive studies.} 
  \item[b.] \textbf{AI climate emulators \textcolor{black}{might} mischaracterize extreme weather statistics \textcolor{black}{for some types of events}:} Our results suggest that current AI climate emulators may not be able to reliably reduce sampling error in gray-swan-event statistics, such as return periods. The \textcolor{black}{possible} inability to reduce this type of error, which arises due to internal variability, undermines a major motivation behind developing climate emulators. This is because to do so the emulators would have to learn about gray swan events from the weaker events in the training set, which we find they cannot do \textcolor{black}{for TCs}. As a result, the long, large ensembles of synthetic data these emulators generate \textcolor{black}{may} not contain reliable information about extreme events stronger and rarer than those that exist in the (typically short) training set, and the accuracy of the estimated statistics will be still limited by  the length of the training set. \textcolor{black}{Again, this potential limitation might vary among different types of extreme events.}
  \item[c.] \textbf{Proper definitions/metrics for gray swans and unprecedented rare events for AI models is necessary:} Given the partial ability of FourCastNet to generalize learning dynamically similar \textcolor{black}{storms} across regions as well as differences in common normalization schemes in climate analysis and AI algorithms, the question of which events qualify as gray swans needs to be carefully examined. \textcolor{black}{First, an event considered unprecedented in one region may already have occurred elsewhere. Second,} in climate science, unprecedented events are often defined based on comparing the anomaly of one variable (e.g., near-surface temperature or rainfall) to \textit{local} statistics. Current AI weather models often normalize each input variable with respect to its global mean and standard deviation. Our findings suggest that accounting for the presence of dynamically similar events in other regions, examining \textcolor{black}{global} rather than local fields, can be important in determining whether an event is truly out-of-distribution for an AI \textcolor{black}{weather} model \textcolor{black}{or climate emulator}.
\end{enumerate}


There are some potential remedies that could help AI weather and climate models with forecasting and emulating gray swans. Common AI strategies for dealing with data imbalance are using weighted loss functions and re-sampling. Weighted loss functions can improve the learning of rare events that exist in the training set (see refs. \cite{miloshevich2023probabilistic,sun2023quantifying,yang2024overcoming} for applications in climate science). But this approach will not help with gray swans. Up-sampling is a potential remedy, but it requires generating strong synthetic TCs and including them in the noTC training set. Generative AI models (e.g., diffusion models) might help here ~\cite{bracco2024machine}; however, such models have not yet been shown to generate physically realistic extreme events beyond what was in their own training set. In fact, iterative training of generative models on their own output has been shown to lead to model collapse~\cite{shumailov2024ai}. A more promising approach is to use physics-based or theoretical models to generate strong synthetic TCs~\cite{willson2024dcmip2016}. While such TCs would satisfy physical constraints, how to properly include them in snapshots of more realistic global circulation without causing unphysical predictions needs to be investigated. 

Incorporating physical constraints, such as conservation laws and symmetries, is often mentioned as a method of improving the representation of extreme events in AI models~\cite{chattopadhyay2021towards,chattopadhyay2019analog,wang2022data}. While such approaches have shown promise in toy models and idealized systems, their applicability to state-of-the-art models and complex data has not been demonstrated. Furthermore, whether such constraints would help with gray swans remains to be seen. Here we show that gradient-wind balance is not noticeably worse in the predictions of FourCastNet-noTC than in FourCastNet-Full. This suggests that~\textcolor{black}{the primary reason FourCastNet-noTC does not capture Category 5 TCs is not lack of gradient-wind balance but rather the absence of training samples with comparable intensities.} Even if this balance were somehow enforced, which would not be a trivial task, the representation of gray swans may not improve.    

Finally, an emerging remedy would integrate innovations in mathematical methods for rare events with AI weather/climate models. Applying rare-event sampling algorithms to conventional physics-based geophysical models has already shown promise in producing rare extreme events and even gray swans as well as estimating their statistics, such as return periods \citep{ragone2018computation,webber2019practical,ragone2021rare,abbot2021rare,finkel2023revealing,finkel2024bringing}. The re-sampling of physics-based models, for example a GCM, is guided by a quick best guess of which simulations are progressing toward the rare event of interest. Cheap ensemble forecasts with an AI weather model, or an AI model designed for the specific rare event \citep{jacques2022deep,zhang2024using,abbot2024ai}, could be used to provide the quick best guess. Stronger, rarer extreme weather events produced by the physics-based model (e.g., a numerical weather/climate model) using rare-event simulation can then be added to the training set of the AI weather/climate model. \textcolor{black}{A related approach that is also promising is ensemble boosting~\citep{fischer2023storylines}.} 

In addition to remedies, rigorous tests are needed to quantify the fidelity of any novel AI model in predicting/emulating gray swans. The approach used in this paper is unambiguous and also provides a straightforward way to examine the physical consistency of the predicted events. However, this approach is computationally demanding as it requires training the AI model from random weights multiple times. An alternative approach is to use an emulator to generate a synthetic dataset that is much longer than the training set and compare extreme event statistics with those of a proper ground truth. Note that in this approach, the physical consistency of the rare events would need to be fully examined in addition to summary statistics like return periods. A major challenge with this approach is that it requires a ground truth, e.g., a long dataset of at least hundreds of years, which is not available for ERA5, but can be produced with physics-based climate models. 

Our work demonstrates the importance of testing and improving the behavior of AI weather/climate models on out-of-distribution gray swan events in addition to tests on typical global weather with measures such as the ACC, or on extreme weather events from distributions similar to that of the training set. We have shown that out-of-distribution extrapolation is non-trivial for extreme weather events and moving forward, the burden of proof is on anyone who claims it. This should be done with carefully designed metrics and reliable ground truths, as weak baselines and reporting biases can lead to overoptimism and misleading results~\cite{mcgreivy2024weak}. Given the outsized societal impact of extreme weather events, this charts a critical path forward for improving and fully exploiting powerful new AI weather and climate models.

\newpage
\section*{Methods and Data}

\subsection*{ERA5 data and its TCs}

All training and testing data are derived from the $0.25$~degree ECMWF's ERA5 reanalysis \cite{hersbach2020era5}, available through the Copernicus Climate Data Store (CDS, \url{https://cds.climate.copernicus.eu/}). As in \cite{pathak2022fourcastnet}, data from 1979 to 2015 are used for training. Testing experiments in this study utilize data from 2018 to 2023. Each Category 5 TC test case consists of a control member (initial condition, IC, directly from ERA5 \textcolor{black}{at 0.25-degree resolution}) and 50 members with perturbed ICs. The 50 perturbations are obtained from the Ensemble of Data Assimilations (EDA) from ECMWF, accessible via the THORPEX Interactive Grand Global Ensemble (TIGGE) Data Retrieval (\url{https://apps.ecmwf.int/datasets/data/tigge/}). \textcolor{black}{The EDA data, originally at 0.5-degree resolution, were re-gridded to the 0.25-degree grid to match the control member and be usable as FourCastNet's input.} The perturbation fields of the EDA ensemble are first derived as the difference between the EDA members and their mean. We then re-scale these fields by a factor of 0.1 and add them to the ERA5 field to generate 50 members of ICs for testing (inference).

TCs are identified within the ERA5 dataset by tracking closed mslp contours, with the minimum pressure serving as the criterion for intensity. The TC center is identified at the minimum mslp location. For the gradient-wind analysis, the TC center is defined as the location of the lowest 500~hPa geopotential height, given a possible vertical tilt in the TC structure.

\subsection*{Training FourCastNet with 5 training sets}

FourCastNet~\cite{pathak2022fourcastnet} 
is a recently developed global data-driven deep learning-based weather forecasting model that autoregressively predicts $\mathbf{x}(t+\Delta t)$ from $\mathbf{x}(t)$
\begin{equation}
    \mathbf{x}(t+\Delta t)=\mathcal{M}(\mathbf{x}(t),\theta)
\end{equation} 
where $\mathbf{x}(t)$ is the 3D state of the atmosphere consisting of $33$ atmospheric variables from ERA5 and $\theta$ represents the trainable parameters of the model. FourCastNet uses the Adaptive Fourier Neural Operator (AFNO) framework~\cite{li2020fourier,guibas2021adaptive} to efficiently parameterize the attention mechanism in a vision transformer. Both training and testing (inference) use a timestep of $\Delta t=6$~h.  
We use  FourCastNet's official code~\footnote{https://github.com/NVlabs/FourCastNet} and exact same state variables, architecture, and hyperparameters as in Pathak \textit{et al.}~\cite{pathak2022fourcastnet}. The one difference between the training procedure in Pathak \textit{et al.}~\cite{pathak2022fourcastnet} and the one here is that we add zero-mean Gaussian noise with a variance of $0.3$ to the inputs, i.e. $\mathbf{x}(t)$, during training. We find this to be essential for the stability of all generated ensemble forecasts during inference. 

For each training set (e.g., Full, noTC, etc.), we train five models, each starting from different random initializations of the trainable parameters to account for the uncertainty in the generalization error of the model. Each model is first trained for 80 epochs with a cosine learning-rate schedule at a starting learning rate of $\ell_1\!=\!5\times 10^{-4}$ and then it is fine-tuned for an additional 50 epochs using a cosine learning-rate schedule with a lower learning rate of $\ell_2\!=\!10^{-4}$. The full training of one model takes roughly 7 days on 4 A100 GPUs. 

\subsection*{Preparing the 5 training sets}
The histogram of the minimum of mslp in the tropics ($30^\mathrm{o}$S-$30^\mathrm{o}$N) for each sample in the training data is shown in Figure~\ref{fig:schematic}. The plot identifies critical thresholds: the $5$th percentile ($\sim 970.0$ hPa) and the 25th percentile ($989.0$ hPa). The fine-tuning step of FourCastNet requires two future timesteps, thus, in addition to $\mathbf{x}(t)$ (sample for which mslp falls below 25th in tropics), we must remove two additional samples ($\mathbf{x}(t-2\Delta t)$ and $\mathbf{x}(t-\Delta t)$) from the training set. We therefore set the critical mslp threshold to 988.0 hPa to ensure that just 25\% of the training data is excluded. The resulting training set (noTC) is used to train FourCastNet-noTC.
Note that the 25th percentile threshold of mslp ($< 988.0$~hPa) roughly corresponds to pressure at the center of major Category 3-5 TCs within the ERA5 dataset. Based on the IBTrACs data, approximately $24.8\%$ of storms become major TCs (Category 3–5), and the median value of mslp in ERA5 for Category 3 TCs is $987.0$~hPa, fairly close to the $988.0$~hPa threshold we used (\textcolor{black}{see Figure~S5}). We define Category 5 in ERA5 as TCs that have mslp less than 970.0 hPa, sustained for 12 hours or longer. Note that $970.0$~hPa is lower than the average mslp values for Category 5 TCs of IBTraACs in the ERA5. There are in fact more than 20 Category 5 TCs in the real world during 2018-2023 based on the Saffir–Simpson hurricane wind scale (the average number is 5-7 per year). In this study, we train and test the AI model within the world of the ERA5 dataset. We use the Category 1-5 terminology mainly to facilitate communication. 


FourCastNet-Rand is trained with the Rand dataset. Rand incorporates all the excluded samples in the noTC dataset. To match the size of the noTC training set, we randomly remove samples that do not include Category 3-5 TCs. Thus, Rand includes all major TCs and has the same the total number of training samples and the annual cycle distribution as noTC.


The noWP (noNA) training set excludes samples when there are mslp values below 988.0 hPa in the tropical Western Pacific (North Atlantic) basin. There are many more strong TCs in the Western Pacific compared to the North Atlantic basin. This holds true for both the ERA5 data and observations. In our study, $\sim$51\% of TCs removed in noTC are due to cyclones in the Western Pacific, while only $\sim$~9\% of the removed samples are due to TCs in the North Atlantic.


\bibliography{ref}

\section*{Acknowledgements}
\textcolor{black}{We thank the editor and two anonymous reviewers for insightful comments and suggestions.} This work was supported by ONR award N000142012722 (to P.H.), ARO grant W911NF-22-2- 0124 (to D.A. and J.W.), and NSF grant AGS-2046309 (to P.H.). Computational resources were provided by NSF ACCESS (allocation ATM170020), NCAR's CISL (allocation URIC0009), and the University of Chicago Research Computing Center. 

\section*{Data, Materials, and Software Availability}
We use the original FourCastNet with modifications for our customized training sets. These codes are publicly available  at \href{https://github.com/envfluids/FourCastNet}{https://github.com/envfluids/FourCastNet}. The necessary data to reproduce the results, including the weights of the 25 trained models and indices of dates that are removed in each training dataset, can be found on Zenodo at \href{https://zenodo.org/uploads/13835657}{https://zenodo.org/uploads/13835657} and \href{https://zenodo.org/uploads/13834149}{https://zenodo.org/uploads/13834149}.

\section*{Author contributions statement}
P.H., D.A., and J.W. conceived the experiments. Y.S. prepared the training and testing sets; M.Z. and A.C. trained different versions of FourCastNet and conducted inference. Y.S. analyzed the results. Y.S., P.H., and D.A. wrote the paper.  All authors reviewed the manuscript. 

\section*{Competing interests}
P.H. is a member of the NVIDIA's Advisory Council on Physics ML for Climate Science.

\end{doublespace}

\newpage

\renewcommand{\thefigure}{S\arabic{figure}}
\renewcommand{\thetable}{S\arabic{table}}
\renewcommand{\theequation}{S\arabic{equation}}
\setcounter{figure}{0}
\setcounter{table}{0}
\setcounter{equation}{0}


\nolinenumbers  

\vspace{50mm}
\section*{Supporting Information for \\ ``Can AI weather models predict out-of-distribution gray swan tropical cyclones?''}


\vspace{15mm}

\author{
Y. Qiang Sun\textsuperscript{1,*},
Pedram Hassanzadeh\textsuperscript{1,2*},
Mohsen Zand\textsuperscript{3},
Ashesh Chattopadhyay\textsuperscript{4}
Jonathan Weare\textsuperscript{5}
Dorian S. Abbot\textsuperscript{1}
}

\vspace{10mm}
\noindent\textsuperscript{1}University of Chicago, Department of the Geophysical Sciences, Chicago, IL 60637, USA\\
\textsuperscript{2}University of Chicago, Committee on Computational and Applied Mathematics, Chicago, IL 60637, USA\\
\textsuperscript{3}University of Chicago, Research Computing Center, Chicago, IL 60637, USA\\
\textsuperscript{4}University of California, Department of Applied Mathematics, Santa Cruz, CA 95064, USA\\
\textsuperscript{5}New York University, Courant Institute of Mathematical Sciences, New York, NY 10012, USA\\

\textsuperscript{*}Corresponding authors: qiangsun@uchicago.edu and pedramh@uchicago.edu

\vspace{15mm}
\subsection*{Content of this file}

\vspace{5mm}
\begin{enumerate}
\item Figures S1 to S5
\item Table S1
\end{enumerate}

\newpage
\begin{figure}[htbp]
\centering
\includegraphics[width=0.95\textwidth]{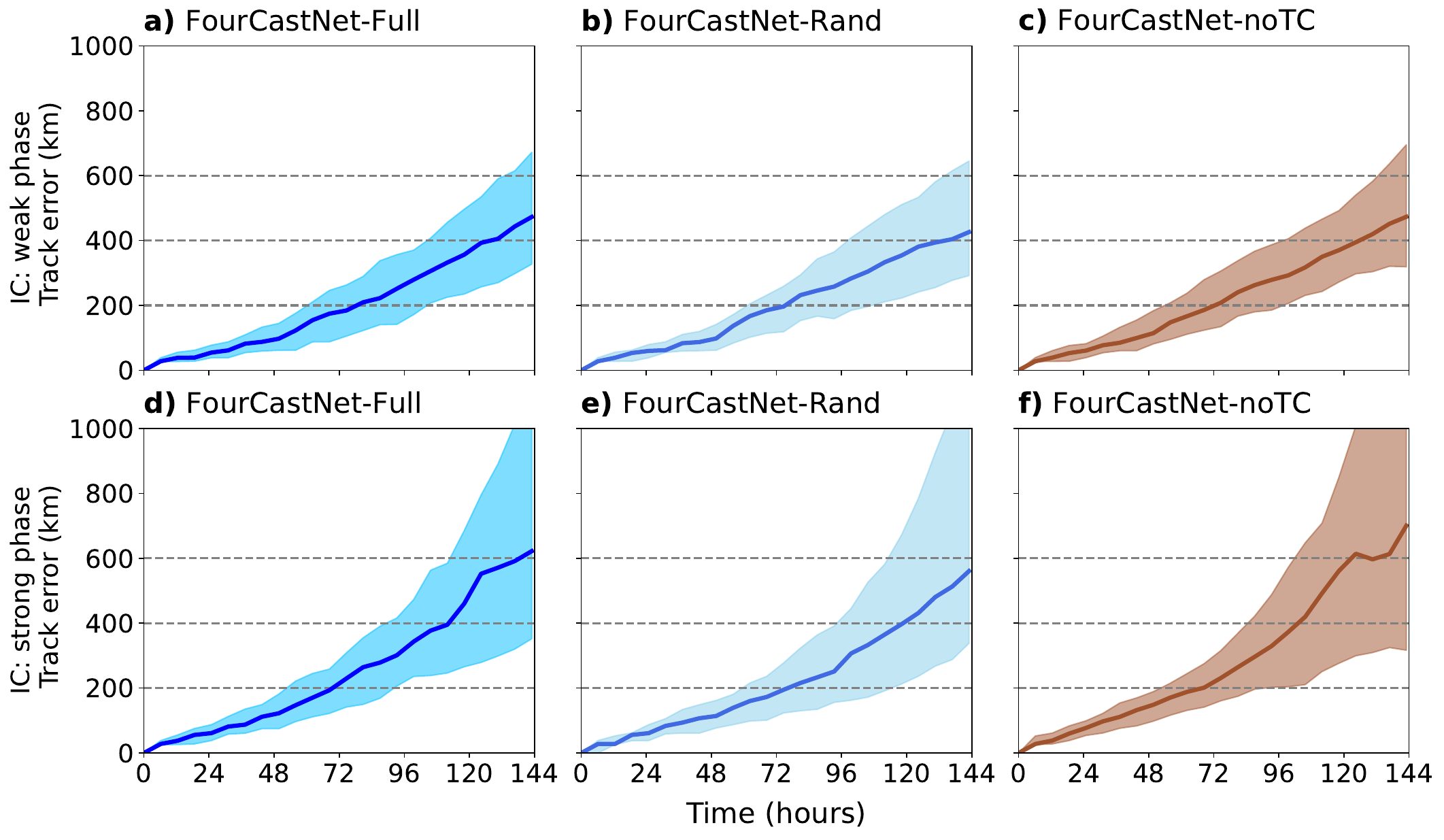}
\caption{{\textbf{All versions of FourCastNet show similar performance in predicting TC tracks}. The forecast errors of the tracks for Category 5 TCs in the test data are shown with shading representing the 25th and 75th percentiles across all forecasts. All forecast data are initialized as in Figure 2. Forecasts with each lead time is generated using 5 realizations and 51 initial conditions (ICs) on all 20 TCs, as described in the main text. Therefore, for each lead time, we have 5 × 51 × 20 = 5100 values. TC track error is defined as the distance between the location of the TC center (minimum mslp) in each forecast and the actual location of the TC in the ERA5 data.}} 
\label{fig:SItrack}
\end{figure}

\begin{figure}[htbp]
\centering

\includegraphics[width=1.05\textwidth]{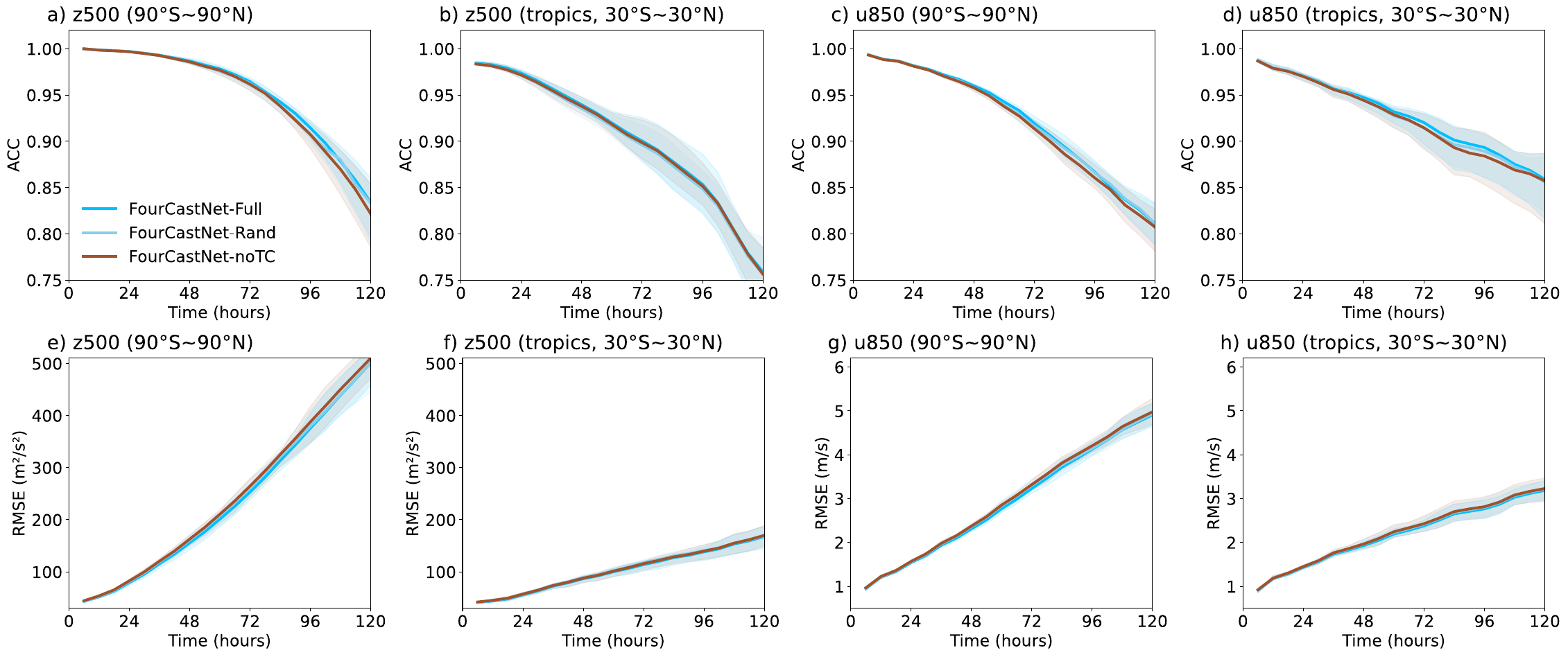}
\caption{{\textbf{All versions of FourCastNet show similar forecast skill when evaluated with metrics that focus on large-scale weather, such as ACC ((a)-(d)) and RMSE ((e)-(h))}. \textbf{(a)} ACC for geopotential at 500~hPa (z500) over the globe. \textbf{(b)} ACC for z500 over the tropics (30\degree S - 30\degree N) only. \textbf{(c)} ACC for zonal wind at 850~hPa (u850) over the globe. \textbf{(d)} ACC for u850 over the tropics (30\degree S - 30\degree N) only. \textbf{(e)-(h)} Similar to \textbf{(a)-(d)}, but for RMSE instead of ACC. We define the ACC and RMSE as the mean latitude-weighted value over
all forecasts, following WeatherBench~\cite{rasp2020weatherbench}. The shading represents the 25th and 75th percentiles across all the forecasts. Forecasts are initialized on the 1st and 15th of each month in 2018 for each trained version of FourCastNet. Each forecast is generated using 5 realizations and 51 initial conditions (ICs), as described in the main text.}}
\label{fig:SIacc}
\end{figure}

\begin{figure}[htbp]
\centering
\includegraphics[width=0.7\textwidth]{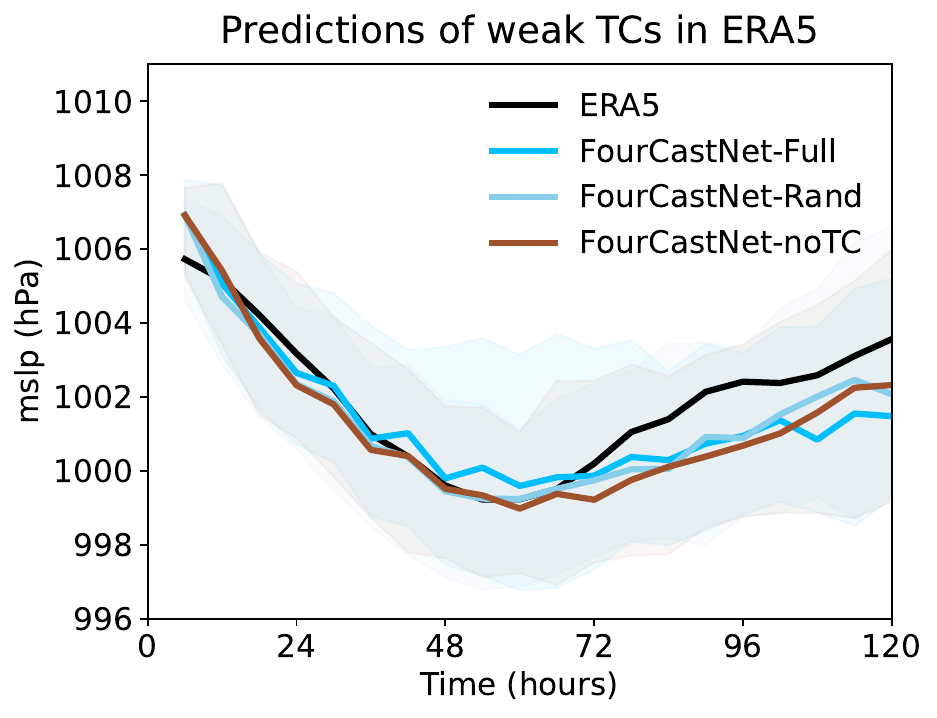}
\caption{{\textbf{All versions of FourCastNet show similar performance on weaker TCs.} Evolutions of mslp at the center of Category 1-2 TCs in ERA5. The lines represent the median values in ERA5 (black line) and different versions of FourCastNet's forecasts (colored lines). Two criteria are used to define weaker TCs: a) TCs are classified as Category 1 or 2 in the IBTrACS data; b) the TC center reached a minimum mslp between $988.0$ hPa and $1000.0$ hPa in the ERA5 data during their life cycle. All forecasts are initialized 2 days before each TC reaches Category 1 in the IBTrACS data. The shading represents the 25th and 75th percentiles of the predicted values.}}
\end{figure}

\begin{figure}[htbp]
\centering
\includegraphics[width=0.65\textwidth]{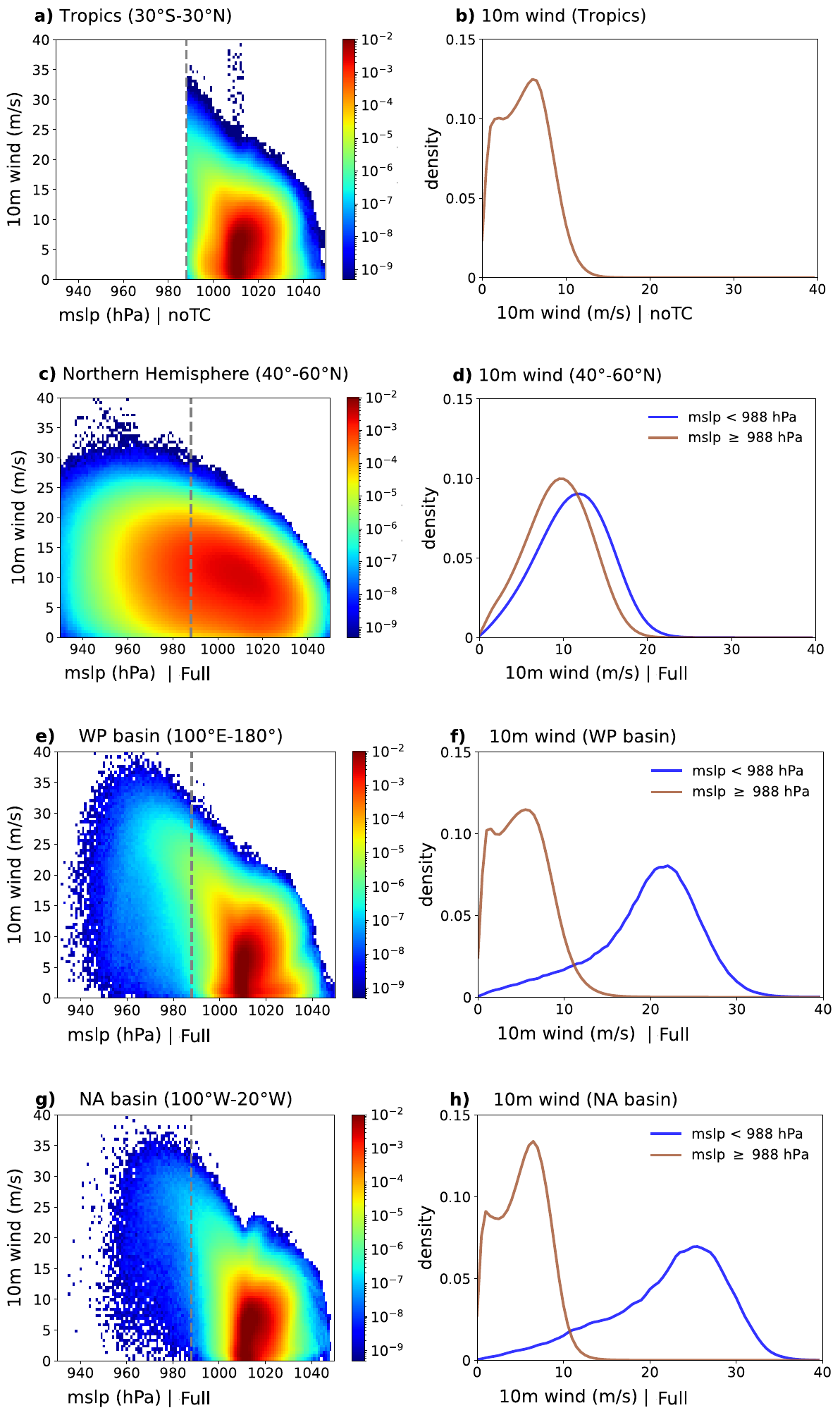}
\caption{{\textbf{Dynamical similarity between the North Atlantic (NA) and Western Pacific (WP) tropical basins.} \textbf{a)} Similar to Figure 3(a) in the main text, except that here we show the joint PDF between 10-m winds and mslp in the noTC dataset. As expected, mslp is always larger than 988.0~hPa. \textbf{b)} Similar to Figure 3(b) in the main text, but for the noTC dataset. \textbf{c)} Similar to Figure 3(c) in the main text, except that here we show the joint PDF between 10-m winds and mslp of the mid-latitudes in the Full dataset. \textbf{b)} Similar to Figure 3(d) in the main text, but for the Full dataset.
\textbf{e)} Similar to Figure 3(a) in the main text, but for the WP basin (box region between $0\degree - 30\degree$~N and $100\degree$~E$\, -\, 180\degree$ ). \textbf{f)} Similar to Figure 3(b) in the main text, but for the WP basin. \textbf{g)} Similar to Figure 3(c) in the main text, but for the NA basin (box region between $0\degree - 30\degree$~N and $100\degree$~W$\, -\, 20\degree$~W ). \textbf{h)} Similar to Figure 3(d) in the main text, but for the NA basin.}}
\end{figure}

\begin{figure}[htbp]
\centering
\includegraphics[width=0.8\textwidth]{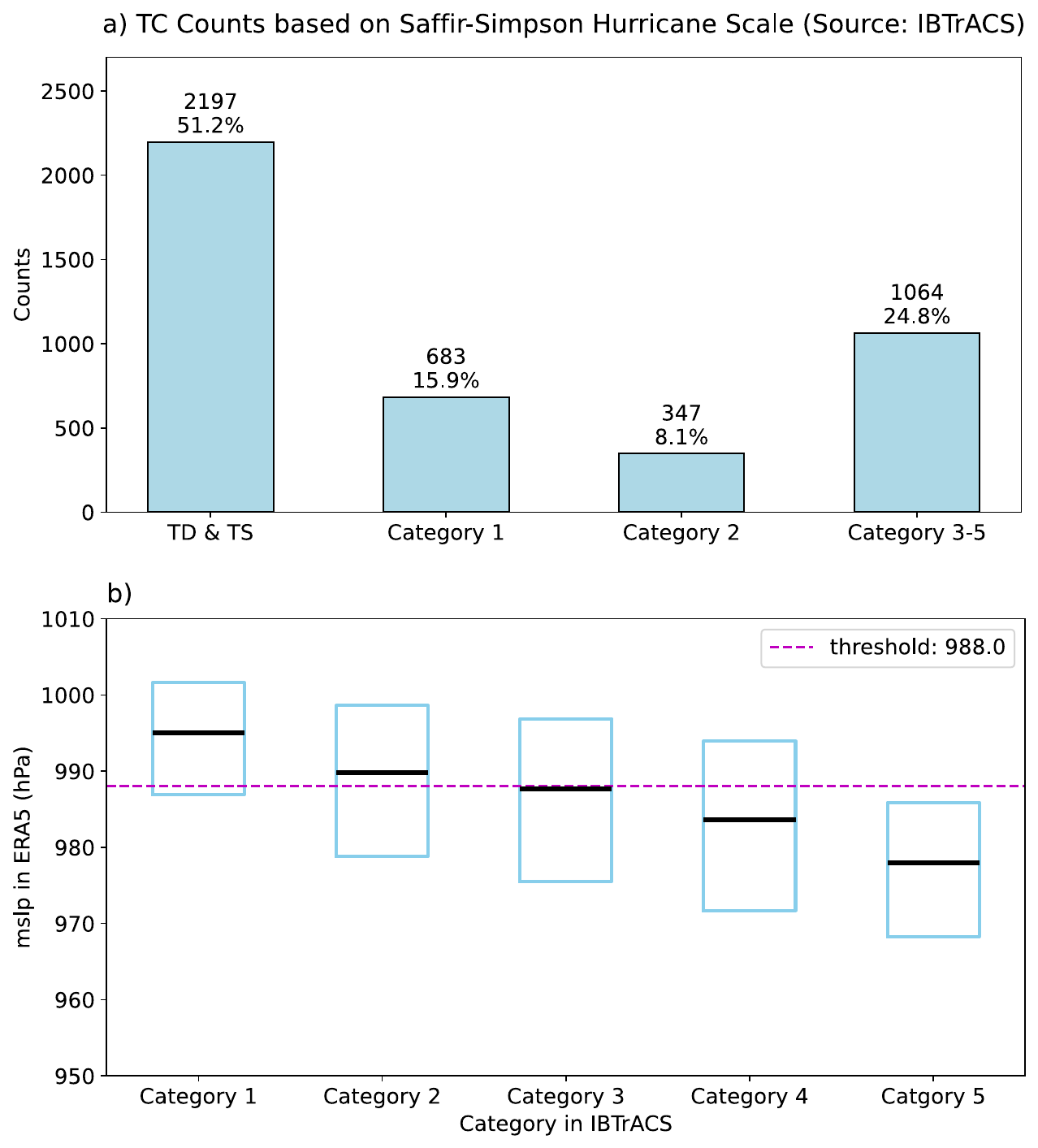}
\caption{ \textbf{TC categories based on IBTrACS data and their corresponding mslp values in ERA5.} \textbf{a)} TC counts for each category based on IBTrACS data, showing that approximately 24.8\% of storms develop into major TCs (Category 3–5). \textbf{b)} The corresponding ERA5 mslp values for each TC category (based on IBTrACS data). The black line represents the median value for each category, with the top and bottom of the box indicating the 25th and 75th percentiles of the ERA5 mslps for that category. For each TC, a time series of ERA5 mslp values at the storm's center is tracked throughout its life cycle.  Note that all real-world Category 1–5 TCs passed through the Category 1 stage, so their ERA5 mslp values at that stage are included in the Category 1 bar plot. The same applies to the bar plots for Categories 2–5.}
\end{figure}

\vspace{20mm}

\newpage
\begin{table}[htbp]
\centering
\begin{tabular}{lrrr}
\textbf{TC name} & \textbf{Basin} & \textbf{yyyy/mm/dd} \\
\hline
1.  AMPHAN & Northern Indian Ocean &  2020-05-15  \\
2.  DUMAZILE & Southern Indian Ocean & 2018-03-03  \\
3.  HAGIBIS & WP & 2019-10-05 \\
4.  HAISHEN & WP & 2020-09-01  \\
5.  JEBI & WP & 2018-08-29  \\
6. KHANUN & WP & 2023-07-28 \\
7. KONGREY & WP & 2018-09-29 \\
8. LARRY & NA & 2021-09-03 \\
9. LEE & NA & 2023-09-07 \\
10. LORENZO & NA & 2019-09-25 \\
11. MANGKHUT & WP & 2018-09-09 \\
12. MAWAR & WP & 2023-05-22 \\
13. MAYSAK & WP & 2020-08-28 \\
14. MINDULLE & WP & 2021-09-24 \\
15. NANMADOL & WP & 2022-09-13 \\
16. SURIGAE & WP & 2021-04-14 \\
17. TEDDY & NA & 2020-09-16 \\
18. TRAMI & WP & 2018-09-22 \\
19. YUTU & WP & 2018-10-22 \\
20. ERA5 & Southern Hemisphere & 2018-03-28\\
\hline
\vspace{10pt}
\end{tabular}
\caption{List of all 20 Category-5 TCs in the ERA5 test dataset from 2018 to 2023, and their corrsponding names from IBTrack. Of the 20, 13 occurred in the Western Pacific (WP), 4 in the North Atlantic (NA), 1 in the Northern Indian Ocean, and 2 in the Southern Indian Ocean. The last TC exists only in the ERA5 dataset, having been generated by the IFS model (model error) and lacks a corresponding observed TC name.}
\end{table}

\end{document}